\documentstyle[preprint,eqsecnum,aps]{revtex}
\tightenlines
\begin{document}
\draft
\title{Comment on ``Indication, from Pioneer 10/11, Galileo and Ulysses 
Data, of an Apparent Anomalous, Weak, Long-Range Acceleration''} 
\author{J. I. Katz}
\address{Department of Physics and McDonnell Center for the Space
Sciences\\Washington U., St. Louis, Mo. 63130\\katz@wuphys.wustl.edu}
\date{\today}
\maketitle
\narrowtext
This paper\cite{A98} may have underestimated the acceleration resulting from
the radiation of waste heat by the Pioneer spacecraft RTG.  These generators
are not very efficient; theoretical efficiencies may be as high as
20\%\cite{D97} but RTG used in spacecraft more typically have electrical
efficiencies of about 6\% when new\cite{TOPS}, which decline slowly as the
thermoelectric material degrades and radioactive decay reduces the Carnot
efficiency.  Specific parameters and detailed design information for Pioneer
are difficult to obtain so I adopt an initial electrical efficiency of 6\%.
Then the electric power at launch of 160 W\cite{A98} implied a thermal power
of 2.67 kW and a waste heat of 2.51 kW.  In 1997 the thermal power, decaying
with the half life of Pu$^{238}$ of 87.74 y, was 2.19 kW and the electrical
power of 80 W\cite{A98} implied a waste heat of 2.11 kW.

The power of a collimated beam sufficient to explain the reported anomalous
acceleration $a_P$ is 85 W\cite{A98}.  The same force can be obtained from
the present rejected waste heat if it is radiated with $\langle \cos\theta
\rangle = 0.040$, where $\theta$ is the angle between the direction of
radiation and a ray from the Sun.  If the 72 W of dissipated electrical
power (allowing for 8 W radiated by the antenna) are radiated from the back
of the spacecraft according to Lambert's law the RTG waste heat need only
give the thrust of a 37 W collimated beam, requiring $\langle \cos\theta
\rangle = 0.018$.  Any components of recoil force orthogonal to the spin
axis are averaged to zero by the spin.  This axis and the net force point
towards the Earth, and at the present great distances this closely aligns
them with the Sun; the Solar alignment is almost exact after averaging over
Earth's orbit.

The Pioneer spacecraft\cite{H74} have their RTG mounted on booms somewhat
behind their high gain antennae.  Thermal radiation scattered from the back
of the antenna will be preferentially directed away from the Sun, leading to
a small postive $\langle \cos\theta \rangle$.  Engineering drawings\cite{PPD}
determine the geometry sufficiently to permit a calculation of $\langle
\cos\theta \rangle$ by numerical integration.  Scattering by spacecraft
components other than the antenna dish is ignored and the RTG are assumed to
radiate as isotropic point sources at their centers.  I find $\langle
\cos\theta \rangle = 0.018$ if the back of the dish is assumed to be a
specular reflector and $\langle \cos\theta \rangle = 0.022$ if it is assumed
to scatter according to Lambert's law.  These results satisfactorily explain
the anomalous acceleration if most of the dissipated electrical power is
radiated from the back of the spacecraft.

This argument may also be applied to the $a_P$ measured\cite{A98} for
Ulysses.  Its design\cite{S92} resembles that of Pioneer, and is generic
to RTG-powered outer Solar System missions.  There is no general reason why
the Pioneer and Ulysses $a_P$ should be approximately equal (the value $a_P
= (12 \pm 3) \times 10^{-8}$ cm/s$^2$ for Ulysses is uncertain enough that
the equality need only be to a factor of two), but they will be similar for
spacecraft of similar geometric design and power to mass ratio.

Anderson, {\it et al.}\cite{A98} point out that the $a_P$ they measure for
Pioneer are constant to within $2 \times 10^{-8}$ cm/s$^2$, about 24\% of
the mean $a_P$, over the range 40--60 AU (about 9 years).  The accuracy of
their quoted values for $a_P$ ($(8.09 \pm 0.20) \times 10^{-8}$ cm/s$^2$ for
Pioneer 10 and $(8.56 \pm 0.15) \times 10^{-8}$ cm/s$^2$ for Pioneer 11)
suggests a typographical error, and that constancy may actually be known to
within $2 \times 10^{-9}$ cm/s$^2$, about 2.4\%.  The total thermal power
decayed by about 7\% over this period, but the decreasing electrical
efficiency implies that the waste heat rejected declined by about 6\%, a
number weakly dependent on the assumed initial efficiency.  This is
marginally inconsistent ($2.5\sigma$) with the more stringent estimate of
constancy of the acceleration (effects such as a temperature dependence of
$\langle \cos\theta \rangle$ resulting from frequency-dependent radiative
properties may perhaps remove any discrepancy), but is completely consistent
with the 24\% bound on constancy of $a_P$ actually quoted\cite{A98}.

I thank V. L. Teplitz for discussions and \cite{PPD} and the NSF for support.

\end{document}